\documentclass[showpacs,showkeys,superscriptaddress,aps]{revtex4}
\begin{document}
\begin{flushright}
TIT/HEP-503 \\
July,2003 \\
\end{flushright}
\title{Soliton equations solved by the boundary CFT}
\author{Satoru SAITO}
\email[email : ]{saito@phys.metro-u.ac.jp}
\affiliation{Department of Physics, Tokyo Metropolitan University,\\
Minamiohsawa 1-1, Hachiohji, Tokyo 192-0397 Japan}
\author{Ryuichi SATO}
\email[email : ]{ryu@th.phys.titech.ac.jp}
\affiliation{Department of Physics, Tokyo Institute of Technology,\\
 Tokyo, 152-8551 Japan}
\begin{abstract}
Soliton equations are derived which characterize the boundary CFT {\it a la} Callan {\it et al}. Soliton fields of classical soliton equations are shown to appear as a neutral bound state of a pair of soliton fields of BCFT. One soliton amplitude under the influence of the boundary is calculated explicitly and is shown that it is frozen at the Dirichlet limit.
\end{abstract}
\maketitle
\section{Introduction}

An interesting exact solution of the boundary CFT was found by Callan {\it et al} in \cite{CKLM}. The solution, which minimizes the boundary action, admits a soliton to propagate from the boundary. 

On the other hand it has been shown in \cite{SS} that an arbitrary open string correlation function satisfies the Hirota bilinear difference equation (HBDE)\cite{Hirota,Miwa}, a discrete soliton equation which generates equations of the KP hierarchy in various continuous limits. The purpose of this note is to point out that the boundary CFT solution of \cite{CKLM} satisfies a simple generalization of HBDE and clarify the correspondence of solitons in two theories.

Let us first recall the exact solution of the boundary CFT derived by Callan et al\cite{CKLM}. They consider a scalar field $X(z)$ which has a periodicity $X\sim X+2\pi\sqrt 2$ on the boundary. The SU(2) currents are introduced by
$$
J^\pm(z)=:e^{\pm i\sqrt2 X(z)}:,\qquad J^3(z)={i\over\sqrt 2}{\partial X(z)\over\partial z}
$$
and the interacting boundary state is obtained from the Neumann boundary state $|N\rangle$ by an SU(2) rotation according to the following formula
\begin{equation}
|B\rangle=e^{i\theta_aJ^a}|N\rangle,\qquad \theta_aJ^a=\pi(gJ^++\bar gJ^-)
\label{B-rotation N}
\end{equation}
where $g$ is a complex coupling constant and
$$
J^\pm=\oint{dz\over 2\pi i}J^\pm(z),\qquad J^3=\oint{dz\over 2\pi i}J^3(z).
$$

Under this circumstance $1\rightarrow 1$ particle correlation function is calculated as
$$
\langle \partial X(z,\bar z)\bar\partial X(z',\bar z')\rangle=-{\cos 2\pi g\over(z-\bar z')^2}.
$$
The rest of the probability is scattered into charge sectors according to
$$
\langle \bar\partial X(\bar z')e^{\pm i\sqrt2 X(z)}\rangle=\pm{1\over\sqrt 2}{\sin 2\pi g\over (z-\bar z')^2},
$$
which were interpreted as solitons propagating away from the boundary.
Since the solitons, which appear in the boundary CFT, are topological ones, it seems, at first glance, nothing to do with solutions of classical soliton equations, such as the KdV equation, the Toda lattice, the sine-Gordon equation and so on. 

Despite of this natural guess we would like to show, in this paper, that the BCFT solitons and those of the classical equations share a common origin. We will derive a soliton equation which is satisfied by the boundary CFT correlation functions. A soliton solution to the equation turns out to be a neutral bound state of soliton fields of the boundary CFT.

We review briefly the string-soliton correspondence in the case of open strings in \S 2, so that solutions to the classical soliton equations are expressed in terms of the language of the open string theory. In \S 3 the string-soliton correspondence is extended to the closed string theory. The manifestly symmetric conformal bound states are reexamined within our formulation in \S 4 and the soliton equation, which characterizes the BCFT, is derived in \S 5. The 'soliton fields' in two theories will be compared and their relations will be examined in \S 6.
In the final section we will calculate explicitly one soliton solution under the influence of the boundary.

\section{String-soliton correspondence in the case of open strings}

 In order to clarify our assertion it will be most instructive to study the structure of one soliton solution written in the form of $\tau$ function:
\begin{equation}
\tau_{1sol}=1+{a\over z-z'}e^{\xi(t,z)-\xi(t,z')},
\label{1sol}
\end{equation}
$$
\xi(t,z):=-\sum_{n=0}^\infty t_nz^{n}.
$$
To be specific let us consider the case of Toda lattice as an example. If we put
$$
t_1=m,\ t_3=t,\  t_j=0\ \ (j\ne 1,3)
$$
the function $r_m(t)$ defined by
\begin{equation}
e^{-r_m}=1+{d^2\over dt^2}\ln\tau(t,m)
\label{r-tau}
\end{equation}
satisfies the Toda lattice equation
$$
{d^2r_m\over dt^2}=2e^{-r_m}-e^{-r_{m+1}}-e^{-r_{m-1}}.
$$
The complex parameters $z,z'$ determine the properties of the soliton, {\it i.e.},
$$
e^{-r_m}-1=\omega^2\ \cosh^{-2}(\kappa m+\omega t+\delta),
$$
\begin{equation}
\kappa={1\over 2}(z'-z),\ \ \omega={1\over 2}\left(z'^3-z^3\right),\ \ e^{2\delta}={a\over z-z'}.
\label{Toda velocity}
\end{equation}

Now we show that the $\tau$ function of (\ref{1sol}) can be described in terms of the string theory. For this purpose we introduce free fermion fields $\psi(z)$ and $\psi^*(z)$ satisfying the anti-commutation relations
\begin{eqnarray}
\{\psi(z),\ \psi(z')\}&=&\{\psi^*(z),\ \psi^*(z')\}=0,\nonumber\\
\{\psi(z),\ \psi^*(z')\}&=&2\pi i\delta(z-z')
\label{anti-commutation relation}
\end{eqnarray}
and write $\tau_{1sol}$ as
\begin{equation}
\tau_{1sol}=\langle 0|e^{H(t)}e^{a\psi(z)\psi^*(z')}|0\rangle
\label{tau 2}
\end{equation}
where
\begin{equation}
H(t)=\sum_{n=0}^\infty t_n\oint{dz\over 2\pi i}z^{n}\psi^*(z)\psi(z).
\label{Hamiltonian}
\end{equation}
The vacuum state $|0\rangle$ is assumed to satisfy
$$
H(t)|0\rangle=0.
$$
The Grassmann nature of the fields $\psi,\ \psi^*$ and the properties
\begin{eqnarray*}
e^{H(t)}\psi(z)e^{-H(t)}&=&e^{\xi(t,z)}\psi(z),\\
e^{H(t)}\psi^*(z)e^{-H(t)}&=&e^{-\xi(t,z)}\psi^*(z)
\end{eqnarray*}
will show that (\ref{tau 2}) reproduces the $\tau$ of (\ref{1sol}).

One of the keys to the correspondence between the soliton theory and the string theory is the bosonization of the fields $\psi$. It is done by writing $\psi(z)$ in terms of the string coordinate $X^\mu(z)$:
\begin{eqnarray}
X^\mu(z)&=&X^\mu_+(z)+X^\mu_-(z)\nonumber\\
X^\mu_+(z)&=&x^\mu_0-i\sum_{n=1}^\infty {1\over n}\alpha^\mu_{-n}z^{n},\nonumber\\
X^\mu_-(z)&=&-ip^\mu\ln z+i\sum_{n=1}^\infty {1\over n}\alpha^\mu_{n}z^{-n}
\label{holomorphic coordinate}
\end{eqnarray}
We consider only one dimensional space time, for simplicity, hence do not write the space-time index $\mu$ hereafter. Using the standard relations
\begin{equation}
[x, p]=i,\quad [\alpha_m,\ \alpha_n]=m\delta_{m+n},\quad m,n\ne 0
\label{bosonic relations}
\end{equation}
we can show that
\begin{eqnarray*}
\psi(z)&=&\ :e^{iX(z)}:\ \equiv e^{iX_+(z)}e^{iX_-(z)},\\
\psi^*(z)&=&:e^{-iX(z)}:
\end{eqnarray*}
satisfy the anti-commutation relations (\ref{anti-commutation relation}). Correspondingly the vacuum state $|0\rangle$ is annihilated by $p$ and $\alpha_n,\ n<0$. The Hamiltonian $H(t)$ of (\ref{Hamiltonian}) is now given by
\begin{equation}
H(t)=\oint{dz\over 2\pi}\xi(t,z){\partial X(z)\over\partial z}.
\label{bosonic Hamiltonian}
\end{equation}

Another key to the correspondence is the Miwa transformation of the variables:
\begin{equation}
t_0=-\sum_{j=1}^Mk_j\ln z_j,\quad t_n={1\over n}\sum_{j=1}^M k_jz_j^{-n}.
\label{Miwa}
\end{equation}
The maximum number $M$ of the summation can be chosen arbitrarily. The set of new variables $\{k_j, z_j\}$ enable us to describe the Hamiltonian and the function $\tau_{1sol}$ fully by the language of the string theory. In fact an equivalent expression of $\xi(t,z)$ is given by
$$
\xi(t,z)=\sum_{j=1}^M k_j\Delta_-(z_j,z).
$$
Here by $\Delta_\pm(z,z')$ we denote the 2D Green's function $\ln|z-z'|$ corresponding to the following power series expansions
\begin{equation}
\Delta_\pm(z,z'):=\left(\matrix{\ln|z'|\cr \ln|z|\cr}\right)-\sum_{n=1}^\infty{1\over n}\left({z\over z'}\right)^{\pm n}.
\label{Delta}
\end{equation}
Accordingly we have the string representations of the Hamiltonian and the $\tau_{1sol}$ as
\begin{equation}
H(t)=i\sum_{j=1}^Mk_jX_-(z_j),
\label{H(k,z)}
\end{equation}
\begin{eqnarray*}
&&
\tau_{1sol}\\
&=&\langle 0|:e^{i\sum_jk_jX(z_j)}:\exp\left[a:e^{iX(z)}::e^{-iX(z')}:\right]|0\rangle\\
&=&
\langle 0|:e^{i\sum_jk_jX(z_j)}:\left(1+a:e^{iX(z)}::e^{-iX(z')}:\right)|0\rangle.
\end{eqnarray*}

General solutions to the soliton equations belonging to the KP hierarchy are given, in terms of fermion fields, in the form
\begin{equation}
\tau(t)=\langle 0|e^{H(t)}G|0\rangle.
\label{general tau}
\end{equation}
Here $G$ is an element of the universal Grassmannian introduced by M.Sato\cite{Sato}. In the papers \cite{SS,KS} one of the present authors showed that (\ref{general tau}) can be rewritten by using string coordinates and satisfies explicitly the Hirota bilinear difference equation\cite{Hirota,Miwa}. Namely we can write (\ref{general tau}) as
\begin{eqnarray}
\tau&=&\langle 0|:e^{i\sum_{j} k_jX(z_j)}:\theta\left(\zeta-\oint{dX(z)\over 2\pi}w(z)\right)\nonumber\\
&\times&\exp\left[{1\over 2}\oint{dX(z)\over 2\pi}\oint{dX(z')\over 2\pi}\ln{E(z,z')\over z-z'}\right]|0\rangle
\label{prime tau}
\end{eqnarray}
where $w$, $\theta$ and $E(z,z')$ are the first Abel differential, the Riemann theta function and the prime form, respectively, associated to some Riemann surface\cite{Mumford}. In fact the substitution of (\ref{prime tau}) into the HBDE
\begin{eqnarray}
&&{\tau(k_1+1,k_2,k_3)\tau(k_1,k_2+1,k_3+1)\over (z_1-z_2)(z_1-z_3)}\nonumber\\
&+&
{\tau(k_1,k_2+1,k_3)\tau(k_1+1,k_2,k_3+1)\over (z_2-z_1)(z_2-z_3)}\nonumber\\
&+&
{\tau(k_1,k_2,k_3+1)\tau(k_1+1,k_2+1,k_3)\over (z_3-z_2)(z_3-z_1)}=0
\label{HBDE}
\end{eqnarray}
yields Fay's trisecant formula\cite{Fay}. This formula is correct for any choice of three $k_j$'s among $\{k_1,k_2,k_3,\cdots, k_M\}$.

\section{String-soliton correspondence in the case of closed strings}

So far we have used only open strings to represent the $\tau$ functions. In order to study the boundary CFT we have to incorporate the closed string coordinate. Namely we introduce, in addition to (\ref{holomorphic coordinate}), the anti-holomorphic coordinate
\begin{eqnarray*}
\tilde X(\bar z)
&=&
\tilde X_+(\bar z)+\tilde X_-(\bar z),\\
\tilde X_+(\bar z)
&=&
\tilde x_0-i\sum_{n=1}^\infty {1\over n}\tilde\alpha_{-n}\bar z^{n},\\
\tilde X_-(\bar z)&=&-i\tilde p\ln \bar z+i\sum_{n=1}^\infty {1\over n}\tilde\alpha_{n}\bar z^{-n},
\end{eqnarray*}
whose components are constrained by the same relations as (\ref{bosonic relations}) and commute with $X(z)$. The closed string coordinate is then given by
$$
X(z,\bar z)=X(z)+\tilde X(\bar z).
$$
In order to find a fermionic field $\psi(z,\bar z)$ associated with closed strings, first we notice the following identity:
\begin{eqnarray*}
&&:e^{ikX(z,\bar z)}::e^{ik'X(z',\bar z')}:\\
&=& (z-z')^{kk'}(\bar z-\bar z')^{kk'}:e^{ikX(z,\bar z)+ik'X(z',\bar z')}:.
\end{eqnarray*}
When $k=-k'=1/\sqrt2$ and $z'\ne z,\ \bar z'\ne \bar z$, 
\begin{eqnarray*}
&&\left\{:e^{iX(z,\bar z)/\sqrt2}:,\ :e^{-iX(z',\bar z')/\sqrt2}:\right\}\\
&&=
\left({1\over(z-z')^{1/2}(\bar z-\bar z')^{1/2}}+{1\over(z'-z)^{1/2}(\bar z'-\bar z)^{1/2}}\right)\\
&&\qquad\quad\times :e^{i\left(X(z,\bar z)-iX(z',\bar z')\right)/\sqrt2}:\\
&&=0
\end{eqnarray*}
holds. At $z'=z,\ \bar z'=\bar z$ the quantity in the bracket $()$ diverges. Hence we obtain a bosonization for the closed strings
\begin{equation}
\psi(z,\bar z)=:e^{iX(z,\bar z)/\sqrt2}:,\quad 
\psi^*(z,\bar z)=:e^{-iX(z,\bar z)/\sqrt2}:
\label{closed string bosonization}
\end{equation}
which satisfy
$$
\{\psi(z,\bar z),\ \psi(z',\bar z')\}=\{\psi^*(z,\bar z),\ \psi^*(z',\bar z')\}=0,
$$
\begin{equation}
\{\psi(z,\bar z),\ \psi^*(z',\bar z')\}=(2\pi i)^2\delta(z'-z)\delta(\bar z'-\bar z).
\label{psi,psi*}
\end{equation}

To define the vacuum state of the fermionic fields we expand the fields $\psi(z,\bar z)$ and $\psi^*(z,\bar z)$ as
\begin{eqnarray*}
\psi(z,\bar z)
&=&
\sum_{m,n=-\infty}^\infty \psi_{mn}z^m\bar z^n,
\\
\psi^*(z,\bar z)
&=&
\sum_{m,n=-\infty}^\infty \psi^*_{mn}z^{-m-1}\bar z^{-n-1}.
\end{eqnarray*}
Using (\ref{psi,psi*}) we find
\begin{equation}
\{\psi_{mn},\psi^*_{rs}\}=\delta_{mr}\delta_{ns}.
\end{equation}

If we impose for the vacuum state to satisfy
\begin{eqnarray}
\psi_{mn}|0\rangle\ne 0,\quad \langle 0|\psi^*_{mn}\ne 0,\quad &{\rm iff}& m,n\geq 0,
\nonumber\\
\psi^*_{mn}|0\rangle\ne 0,\quad \langle0|\psi_{mn}\ne 0,\quad &{\rm iff}& \cases{m, n< 0\cr m<0, n=0\cr m=0, n<0\cr},
\label{fermion vacuum}
\end{eqnarray}
it is compatible with the vacuum state of the bosonic fields. In fact we can show, from (\ref{closed string bosonization}), that the relations
\begin{eqnarray}
&&\lim_{z'\rightarrow z}:e^{iX(z,\bar z)/\sqrt2}::e^{-iX(z',\bar z')/\sqrt2}:
\nonumber
\\
&&=
{i\over\sqrt2}{\partial X(z)\over\partial z}+{i\over\sqrt2}{\partial\tilde X(\bar z)\over\partial\bar z}+{\rm const.}
\label{lim e^Xe^X=X}
\end{eqnarray}
are correct up to some (diverging) constants. Therefore a comparison of the coefficients of the power expansion in both sides yields
\begin{eqnarray*}
\alpha_{\pm n}&=&\sqrt2 \sum_{r,s\in \bf{Z}}\psi_{r,s}\psi^*_{r\pm n,s},\quad n=0,1,2,3,\cdots,
\\
\tilde\alpha_{\pm n}&=&\sqrt2 \sum_{r,s\in \bf{Z}}\psi_{r,s}\psi^*_{r,s\pm n},\quad n=0,1,2,3,\cdots,
\\
&&\alpha_0=p,\qquad \tilde\alpha_0=\tilde p.
\end{eqnarray*}
Hence we obtain
\begin{eqnarray}
\alpha_n|0\rangle&=&\tilde\alpha_n|0\rangle=0,\quad n\ge 0,
\nonumber\\
\langle 0|\alpha_{-n}&=&\langle 0|\tilde\alpha_{-n}=0,\quad n>0,
\label{vacuum nature}
\end{eqnarray}

Now we want to derive a $\tau$ function associated with closed strings. The Grassmannian is defined by means of fermion fields $\psi(z,\bar z)$ and $\psi^*(z,\bar z)$. The Hamiltonian is not obvious in the present case, since a straightforward generalization of (\ref{Hamiltonian}) does not make sense. Instead we generalize (\ref{bosonic Hamiltonian}) to
\begin{eqnarray}
H(t,\tilde t)&=&
\oint {dz\over 2\pi}{\partial X(z)\over\partial z}\xi(t,z)+
\oint {d\bar z\over 2\pi}{\partial\tilde X(\bar z)\over\partial\bar z}\xi(\tilde t,\bar z),\nonumber\\
\label{H(t,bar t)}
\end{eqnarray}
from which follow
\begin{eqnarray}
e^{H(t,\tilde t)}\psi(z,\bar z)e^{-H(t,\tilde t)}&=&e^{(\xi(t,z)+\xi(\tilde t,\bar z))/\sqrt2}\psi(z,\bar z),\nonumber\\
e^{H(t,\tilde t)}\psi^*(z,\bar z)e^{-H(t,\tilde t)}&=&e^{-(\xi(t,z)+\xi(\tilde t,\bar z))/\sqrt2}\psi^*(z,\bar z).\nonumber\\
\label{shifts}
\end{eqnarray}

We then define the $\tau$ function by
\begin{equation}
\tau(t,\tilde t)=\langle 0|e^{H(t,\tilde t)}G|0\rangle.
\label{tau}
\end{equation}

$G$ in (\ref{tau}) is an element of the Grassmannian of the fields $\psi(z,\bar z)$ and $\psi^*(z,\bar z)$. To be specific let $V$ and $V^*$ denote the linear space $\oplus\mbox{\boldmath{C}}\psi(z,\bar z)$ and $\oplus\mbox{\boldmath{C}}\psi^*(z,\bar z)$, respectively. Then $G\in G(V,V^*)$ satisfies
$$
GV=VG,\qquad GV^*=V^*G.
$$
In particular, the following relations must hold:
\begin{eqnarray}
\psi(z,\bar z)G&=&\oint{dz'\over 2\pi i}\oint{d\bar z'\over 2\pi i}A(z,\bar z;z',\bar z')G\psi(z',\bar z'),\nonumber\\
G\psi^*(z,\bar z)&=&\oint{dz'\over 2\pi i}\oint{d\bar z'\over 2\pi i}A(z',\bar z';z,\bar z)\psi^*(z',\bar z')G.\nonumber\\
\label{psi G}
\end{eqnarray}

To see the correspondence between the $\tau$ function (\ref{tau}) and the closed string correlation function, we adopt the Miwa transformations in the form
\begin{eqnarray}
\tilde t_0&=&-\sum_{j=1}^{M}k_j\ln \bar z_j,\\
\tilde t_n&=&{1\over n}\sum_{j=1}^{M}k_j\bar z_j^{-n}, \quad n=1,2,\cdots.
\label{tilde t}
\end{eqnarray}
together with (\ref{Miwa}). The Hamiltonian is expressed in terms of the closed string coordinate $X$
\begin{equation}
H(t,\tilde t)=i\sum_{j=1}^Mk_jX_-(z_j,\bar z_j).
\end{equation}

For a given set of variables
\begin{equation}
K=(k_1,k_2,\cdots,k_M,z_1,z_2,\cdots,z_M,\bar z_1,\cdots,\bar z_{M})
\label{KZ}
\end{equation}
we define the string correlation function $F_G$ with the background $G$ by
\begin{equation}
F_G(K)
=\langle 0|:e^{ik_1X(z_1,\bar z_1)}:\cdots :e^{ik_MX(z_M,\bar z_M)}:G|0\rangle.
\label{F_G}
\end{equation}
The correlation of higher spin particles are obtained from this expression via differentiation with respect to corresonding components of momenta $k_j$'s.

By using the identity
\begin{equation}
{:\phi_1::\phi_2:\cdots :\phi_N:\over \langle 0|:\phi_1::\phi_2:\cdots :\phi_N:|0\rangle}=\ :\phi_1\phi_2\cdots \phi_N:.
\label{::/<>}
\end{equation}
which holds for arbitrary fields $\phi_j$'s, the $\tau$ function (\ref{tau}) is then related to the correlation function by the following formula:
$$
\tau(K)={F_G(K)\over F_1(K)}.
$$

In the case of $G=\exp\left[a\psi(z,\bar z)\psi^*(z',\bar z')\right]$, for instance, we obtain the $\tau$ function describing the one soliton solution:
\begin{eqnarray*}
\tau_{1sol}&=&1+{a\over \sqrt{(z-z')(\bar z-\bar z')}}e^{\Xi(t,\tilde t,z,z')}
\label{one closed soliton}
\end{eqnarray*}
where
\begin{equation}
\Xi(t,\tilde t,z,z'):={\xi(t,z)-\xi(t,z')\over\sqrt 2}+{\xi(\tilde t,\bar z)-\xi(\tilde t,\bar z')\over\sqrt2}.
\label{Xi}
\end{equation}

\section{Conformally symmetric boundary states}

We now turn to the problem of the conformally symmetric boundary states. In this section we study manifestly symmetric conformal boundary states. For this purpose we introduce the following notation
\begin{eqnarray*}
\Phi_\pm&:=&\oint_B{dz\over 2\pi i}\oint_B{dz'\over 2\pi i}{\partial\tilde X(1/z)\over\partial z} \Delta_\pm(z,z'){\partial X(z')\over\partial z'}.
\end{eqnarray*}
Here $\Delta_\pm(z,z')$ are those defined in (\ref{Delta}), and the integrations are taken along the boundary $B$. We notice that they can be also expressed as integrations along certain space time closed paths as
\begin{eqnarray*}
\Phi_\pm&:=&\oint{d\tilde X(1/z)\over 2\pi i}\oint{dX(z')\over 2\pi i}\Delta_\pm(z,z').
\end{eqnarray*}

The Neumann boudary states are defined by
\begin{equation}
|N\rangle=e^{\Phi_+}|0\rangle,\quad \langle N|=\langle 0|e^{\Phi_-},
\label{|Nrangle}
\end{equation}
which are manifestly symmetric under the conformal transformations.

Owing to the relations
\begin{eqnarray*}
\left[X(z), {\partial X(z')\over\partial z'}\right]
&=&
2\pi i\delta(z-z'),
\\
\left[\tilde X(z), {\partial \tilde X(z')\over\partial z'}\right]
&=&
2\pi i\delta(z-z')
\end{eqnarray*}
we can verify
\begin{eqnarray*}
[\Phi_\pm, X(y)]
&=&
\left(\matrix{-\tilde X_+(1/y)+i\tilde p\ln|z|+\tilde x_0\cr \tilde X_-(1/y)\cr}\right)
\end{eqnarray*}
\begin{eqnarray*}
[\Phi_\pm, \tilde X(y)]
&=&
\left(\matrix{-X_+(1/y)+ip\ln|y|+x_0\cr X_-(1/y)-ip\ln|yz'| \cr}\right)
\end{eqnarray*}
$$
[\Phi_\pm,[\Phi_\pm,X(y)] ]=[\Phi_\pm,[\Phi_\pm,\tilde X(y)] ]=0.
$$
We are thus convinced the following relations
\begin{equation}
X(y)|N\rangle=\tilde X(1/y)|N\rangle,\quad \langle N|X(y)=\langle N|\tilde X(1/y)
\label{X|Nrangle=tilde X|Nrangle}
\end{equation}

The meaning of the formulae of (\ref{X|Nrangle=tilde X|Nrangle}) is that when a right moving field reflects at the boundary it turns to a left moving one. In particular we have
$$
:e^{\pm iX(z,\bar z)/\sqrt2}:|N\rangle=:e^{\pm i\sqrt2 X(z)}:|N\rangle
=J^\pm(z)|N\rangle
$$
when $\bar z z=1$. The result by Callan {\it et al} \cite{CKLM}, that the dynamical boundary state $|B\rangle$ is obtained from $|N\rangle$ by the SU(2) rotation, owes to this fact.

Following to the argument of \cite{CKLM} let us further consider the scattering of a field $\partial\tilde X/\partial\bar z$ at the boundary. It is more convenient to introduce $\tilde J^3(\bar z)$ by
\begin{equation}
\tilde J^3(\bar z)={i\over\sqrt2}{\partial \tilde X(\bar z)\over\partial \bar z}.
\label{tilde J^3}
\end{equation}
Since $\tilde J^3$ commutes with $J^\pm$ it hits directly to the Neumann boundary state $\langle N|$ when it acts on $\langle B|$. At the boundary it turns to a holomorphic operator $J^3(z)$ and then undergoes an SU(2) rotation before it is reflected. Namely 
\begin{eqnarray}
\langle B|\tilde J^3(\bar z)&=&
\langle N|J^3(z)e^{-i\theta_aJ^a}\nonumber\\
&=&\langle B|J^{\theta}(z)
\label{3-theta}
\end{eqnarray}
where
\begin{eqnarray}
J^\theta(z)&:=&e^{i\theta_aJ^a}J^3(z)e^{-i\theta_aJ^a}\nonumber\\
&=&\cos(2\pi|g|)J^3(z)+\sin(2\pi|g|){gJ^+(z)-\bar gJ^-(z)\over 2i|g|}.\nonumber\\
\label{J^theta}
\end{eqnarray}

Apart from the particle sector represented by the holomorphic field $\partial X/\partial z$ there appear other fields $\exp[\pm i\sqrt2X(z)]$ if $g\ne 0$. The new sectors created by the reflection were interpreted as a production of charged solitons in \cite{CKLM}.

\section{Soliton equations satisfied by closed string correlation functions}

The purpose of this section is to derive a soliton equation satisfied by the closed string correlation functions with conformally symmetric boundaries. It will be done by generalizing the correlation function (\ref{F_G}) to the case with the boundary:
\begin{equation}
F_G^B(K)
=
\langle B|:e^{ik_1X(z_1,\bar z_1)}:\cdots :e^{ik_MX(z_M,\bar z_M)}:G|0\rangle.
\label{F_G^B}
\end{equation}
This is also a generalization of the formulation of \cite{CKLM} to the case with a background specified by the universal Grassmannian $G$.

We want to derive a bilinear relation satisfied by the correlation function (\ref{F_G^B}). To this end we first consider the following integration:
$$
I=\oint_0{dz\over 2\pi i}\oint_\infty{d\bar z\over 2\pi i}\psi^*(z,\bar z)G|0\rangle\otimes\psi(z,\bar z)G|0\rangle,
$$
where we assume that the contours encircle around $z=0$ and $\bar z=\infty$. Applying the relations (\ref{psi G}) this is equivalent to 
$$
=\oint_0{dz\over 2\pi i}\oint_\infty{d\bar z\over 2\pi i}G\psi^*(z,\bar z)|0\rangle\otimes G\psi(z,\bar z)|0\rangle,
$$
which turns out to be zero, {\it i.e.},
\begin{equation}
I=0,
\label{I=0}
\end{equation}
owing to the fact that $\psi^*(z,\bar z)|0\rangle$ and $\psi(z,\bar z)|0\rangle$ are orthogonal. 

This fundamental property will be verified by rewriting $I$ as
$$
I=\sum_{p,q\in \bf{Z}}G\psi^*_{p,q}|0\rangle\otimes G\psi_{p,q}|0\rangle
$$
Each term of the summation is identically zero due to the nature of the vacuum state (\ref{fermion vacuum}).

We can interprete the formula (\ref{I=0}) as a bilinear sum rule for the correlation functions (\ref{F_G^B}) if we multiply to $I$ the state
\begin{eqnarray*}
&&\langle B|:e^{ik_1X(z_1,\bar z_1)}::e^{ik_2X(z_2,\bar z_2)}:\cdots :e^{ik_MX(z_M,\bar z_M)}:\nonumber\\
&\otimes&
\langle B|:e^{ik'_1X(z_1,\bar z_1)}::e^{ik'_2X(z_2,\bar z_2)}:\cdots :e^{ik'_MX(z_M,\bar z_M)}:
\end{eqnarray*}
{\it i.e.},
\begin{equation}
\oint_0{dz\over 2\pi i}\oint_\infty{d\bar z\over 2\pi i}
F_{\psi^*(z,\bar z)G}^B(K)F_{\psi(z,\bar z)G}^B(K')=0.
\label{F^BF^B=0}
\end{equation}
This is a formula every closed string correlation function must satisfy.

In order to find more useful informations we divide (\ref{F^BF^B=0}) by $F^B_1(K)F^B_1(K')$ and rewrite it as
\begin{eqnarray*}
&&\oint_0{dz\over 2\pi i}\oint_\infty{d\bar z\over 2\pi i}
\langle B|:e^{i\sum_jk_jX(z_j,\bar z_j)}:\psi^*(z,\bar z) G|0\rangle\\
&&\quad \times \langle B|:e^{i\sum_jk'_jX(z_j,\bar z_j)}:\psi(z,\bar z) G|0\rangle=0,
\end{eqnarray*}
where we used (\ref{::/<>}).

Now let us recall the expressions (\ref{closed string bosonization}) of $\psi$ and $\psi^*$ and shift their plus components $X_+(z,\bar z)$ to the left to obtain
\begin{eqnarray}
&&\oint_0{dz\over 2\pi i}\oint_\infty{d\bar z\over 2\pi i}\prod_{j=1}^M\left((z_j-z)(\bar z_j-\bar z)\right)^{(k'_j-k_j)/\sqrt2}
\nonumber\\
&&\times
\langle B|:e^{i\sum_jk_jX(z_j,\bar z_j)-iX(z,\bar z)/\sqrt 2}:G|0\rangle\nonumber\\
&&\times
\langle B|:e^{i\sum_jk'_jX(z_j,\bar z_j)+iX(z,\bar z)/\sqrt 2}:G|0\rangle=0.
\label{prod z-z_j}
\end{eqnarray}

So far we have not specified $k_j$'s and $k'_j$'s. Since they are arbitrary we can choose
\begin{equation}
k'_j=k_j-\sqrt 2\quad (j=1,2,3),\quad k'_j=k_j\quad (j\ne 1,2,3)
\label{k,k'}
\end{equation}
such that the factor 
$$
\prod_{j=1}^M\left((z_j-z)(\bar z_j-\bar z)\right)^{(k'_j-k_j)/\sqrt2}
$$
in (\ref{prod z-z_j}) becomes simply
\begin{equation}
\prod_{j=1}^3{1\over (z-z_j)(\bar z-\bar z_j)}.
\label{poles}
\end{equation}

We are ready to move the contours of integrations in (\ref{prod z-z_j}) toward the boundary. We assume that all of the simple poles of (\ref{poles}) on the $z$ plane are in between the boundary and the origins and those on the $\bar z$ plane are in between the boundary and $\infty$. We notice that, while the contours are moved toward the boundary, they are only singularities which contribute to the integrations. Therefore we obtain the bilinear relation for the $\tau$ function with boundary, which is defined by
$$
\tau^B(k_1,k_2,k_3)={F_G^B(K)\over F_1^B(K)},
$$
as follows:
\begin{eqnarray}
&&
{\tau^B(k_1+\sqrt2,k_2,k_3)\tau^B(k_1,k_2+\sqrt2,k_3+\sqrt2)\over(z_1-z_2)(z_1-z_3)}\nonumber\\
&+&
{\tau^B(k_1,k_2+\sqrt2,k_3)\tau^B(k_1+\sqrt2,k_2,k_3+\sqrt2)\over(z_2-z_1)(z_2-z_3)}\nonumber\\
&+&
{\tau^B(k_1,k_2,k_3+\sqrt2)\tau^B(k_1+\sqrt2,k_2+\sqrt2,k_3)\over(z_3-z_1)(z_3-z_2)}\nonumber\\
&+&
\oint_B{dz\over 2\pi i}\oint_B{d\bar z\over 2\pi i}\prod_{j=1}^3{1\over (z-z_j)(\bar z-\bar z_j)}\nonumber\\
&&\times
\tau^B\left(k_1,k_2,k_3,{-1\over\sqrt2}\right)\tau^B\left(k'_1,k'_2,k'_3,{1\over\sqrt2}\right)\nonumber\\
&=&0.
\label{bilinear eq with boundary}
\end{eqnarray}
In this expression $k'_j=k_j+1/\sqrt2$.

A few comments are in order. 

1) If some of $z_j$'s and $\bar z_l^{-1}$'s are not in between the origins and the boundary, they do not contribute to the summation of (\ref{bilinear eq with boundary}).

2) If there is no boundary and the contours of the integrations are moved away to infinity the second term of (\ref{bilinear eq with boundary}) does not contribute and we obtain a generalization of HBDE (\ref{HBDE}) to the closed strings.

3) If we are concerned only to deriving bilinear identities of the $\tau$ function we could generalize the Miwa transformation (\ref{tilde t}) to
\begin{eqnarray*}
\tilde t_0&=&-\sum_{j=1}^{M}\tilde k_j\ln \bar z_j,\\
\tilde t_n&=&{1\over n}\sum_{j=1}^{M}\tilde k_j\bar z_j^{-n}, \quad n=1,2,\cdots,
\end{eqnarray*}
such that $\tilde k_j$'s are independent from $k_j$'s.
\section{Correspondence of solitons in two theories}

In the paper \cite{CKLM} by Callan {\it et al} it was argued that the field $X$, which was at a minimum of the potential, is displaced by an integer number of period $2\pi\sqrt 2$ after scattering from the boundary. The shift is visible as a topological soliton which is propagating away from the boundary. The generators, which carry the topological numbers, are identified with the weight one field $e^{\pm i\sqrt2 X(z)}$. The exponents describe kinks with a shift of $2\pi\sqrt2$ between values of $X$ before and after the scattering.

On the other hand we have derived in this paper soliton equations which are satisfied by the correlation functions of closed strings. The question is whether solitons which appear in two theories have some relation? The purpose of this section is to clarify this point.

In the theory of KP-hierarchy\cite{DJKM} a creation of one soliton state is generated by the operation of
\begin{equation}
e^{a\Lambda(z)\Lambda^*(z')},
\label{LambdaLambda}
\end{equation}
where
$$
\Lambda(z)=e^{\xi(t,z)}e^{\xi(\partial_t,z^{-1})},\quad \Lambda^*(z)=e^{-\xi(t,z)}e^{-\xi(\partial_t,z^{-1})}
$$
to the $\tau$ function. Here $\partial_t$ means $(\partial_{t_1},\partial_{t_2},\partial_{t_3},\cdots)$. One soliton solution (\ref{tau 2}), for example, will be seen being generated from $\langle 0|e^{H(t)}|0\rangle=1$ by this operation. If we expand $\Lambda(z)\Lambda^*(z')$ as
$$
\Lambda(z)\Lambda^*(z')=\sum_{m,n\in\mbox{\boldmath$Z$}}L_{mn}z^mz'^{-n}
$$
the coefficients satisfy
$$
[L_{mn}, L_{m'n'}]=\delta_{m'n}L_{mn'}-\delta_{mn'}L_{m'n}.
$$
Hence (\ref{LambdaLambda}) is an element of the group $GL(\infty)$.

We can translate this symmetry into a simple word of the language of string theory. Namely the actions of $\Lambda(z)$ and $\Lambda^*(z)$ are equivalent to the shifting $kX(z)$ to $(k\pm 1)X(z)$, which we can express as
$$
\Lambda(z)=e^{\partial_k},\qquad \Lambda^*(z)=e^{-\partial_k}.
$$
If $z$ coincides with one of the $z_j$'s in the Hamiltonian (\ref{H(k,z)}), $\Lambda(z_j)$ increases the corresponding value of $k_j$ by one. If $z$ coincides with none of them, it creates a new field $e^{iX(z)}$ in the string correlation functions. The generation of a soliton in the soliton theory, which is achieved by an action of the operator (\ref{LambdaLambda}), is equivalent to an action of
$$
1+ae^{\partial_k}e^{-\partial_{k'}}
$$
to the string correlation functions. It is apparent that the action of the second term adds a pair of fields $:e^{iX(z)}::e^{-iX(z')}:$ to the correlation functions. Since this process generates new solutions to the HBDE (\ref{HBDE}) it is a B\"acklund transformation \cite{SS2}.

Let us extend above consideration to the closed strings. Instead of $\psi(z),\ \psi^*(z)$ in (\ref{tau 2}) a soliton state will be described by the fields $\psi(z,\bar z)$ and $\psi^*(z,\bar z)$. The bosonization procedure (\ref{closed string bosonization}) enables us to express them in terms of string coordinates. The corresponding generator of the B\"acklund transformation is
$$
\Lambda(z,\bar z)\Lambda^*(z',\bar z')=e^{(1/\sqrt2)\partial_k}e^{-(1/\sqrt2)\partial_{k'}},
$$
which introduces a pair of fields
\begin{equation}
\psi(z,\bar z)\psi^*(z',\bar z')=:e^{iX(z,\bar z)/\sqrt2}::e^{-iX(z',\bar z')/\sqrt2}:
\label{e^iX(z,bar z)/sqrt2e^-iX(z',bar z')/sqrt2}
\end{equation}
into the string correlation functions. We will call this field (\ref{e^iX(z,bar z)/sqrt2e^-iX(z',bar z')/sqrt2}) a 'classical soliton field'.

As it was discussed in \cite{CKLM} the pair of fields of (\ref{e^iX(z,bar z)/sqrt2e^-iX(z',bar z')/sqrt2}) is equivalent to 
$$
:e^{i\sqrt2 X(z)}::e^{-i\sqrt2 X(z')}:
$$
when it is on the boundary. Let us call $e^{\pm i\sqrt2 X(z)}$ the `BCFT soliton fields'. From this observation we find that a pair of `BCFT soliton fields' in \cite{CKLM} form a 'classical soliton field' of classical soliton equations. In other words a soliton of classical equations is a neutral bound state of a pair of charged fields in the boundary CFT.

The profile of the soliton solution to classical equations has been represented by (\ref{1sol}). It is an exponential function of the soliton coordinates $t_n$'s. The special dependence on $t_n$'s implies the solitary nature of the solution when the Toda amplitude $r_m$ is expressed in terms of the $\tau$ function via (\ref{r-tau}). The question is what is the meaning of the classical {\it soliton coordinates} $t_n$ in the string theory?

In order to answer this question we rewrite the Hamiltonian $H(t,\tilde t)$ of (\ref{H(t,bar t)}), which generates the motion of classical solitons, as
\begin{eqnarray}
H(t,\tilde t)&=&
i\sum_jk_jX_-(z_j,\bar z_j)\nonumber\\
&=&
-\sum_{n=0}^\infty(t_n\alpha_n-\tilde t_n\tilde\alpha_n).
\label{H(tt)=}
\end{eqnarray}

We notice that $\alpha_n$'s and $\tilde \alpha_n$'s are particle modes of a string. Therefore the two expressions in (\ref{H(tt)=}) present two different pictures, {\it i.e.} a string picture and a particle picture. This also provides a physical interpretation of the Miwa transformation (\ref{Miwa}).

The direction of motion of a classical soliton depends on which particle mode of the string is excited. Moreover (\ref{shifts}) shows that the velocity of the classical soliton moving along $t_n$ is determined by the eigenvalue $z^n$ of $\alpha_n$ as it was the case of the Toda lattice (\ref{Toda velocity}).

\section{One soliton solution}

To be specific let us study the correlation function of closed strings (\ref{F_G^B}) with the boundary state $\langle B|$ and $G$ is given by $\exp[a\psi(z,\bar z)\psi^*(z',\bar z')]$. If we restrict the values of $k_j$'s to multiple of $1/\sqrt 2$ it is a correlation function of `BCFT soliton fields'. The presence of $G$ amounts to introduce a `classical soliton field' into the correlation function. But, once the pair is substituted into the function, it is not distinguished from other `BCFT soliton fields'.

In the language of $\tau$ functions we have
\begin{eqnarray*}
&&\tau^B_{1sol}=1+a{F^B_{\psi(z,\bar z)\psi^*(z',\bar z')}(K)\over F_1^B(K)}\\
&=&
1+a
{\langle B|:e^{i\sum_jk_jX(z_j,\bar z_j)}:\psi(z,\bar z)\psi^*(z',\bar z')|0\rangle \over\langle B|:e^{i\sum_jk_jX(z_j,\bar z_j)}:|0\rangle}\\
&=&
1+a{W(t,\tilde t)\over \sqrt{(z-z')(\bar z-\bar z')}}e^{\Xi(t,\tilde t,z,z')}
\end{eqnarray*}
where $\Xi$ is given by (\ref{Xi}) and we used (\ref{shifts}) to obtain the exponential factor. The existence of other `BCFT soliton fields' causes an effect on the `classical soliton fields' $\psi(z,\bar z)\psi^*(z',\bar z')$ to move along the direction determined by the $\xi$'s in this exponential factor. 

Comparing with the one soliton state (\ref{one closed soliton}) with no boundary, there is an extra factor $W$ which now depends on $t$'s and is given by
$$
W=
{\langle B|e^{i\sum_jk_jX_+(z_j,\bar z_j)}e^{i(X_+(z,\bar z)-X_+(z',\bar z'))/\sqrt2}|0\rangle \over\langle B|e^{i\sum_jk_jX_+(z_j,\bar z_j)}|0\rangle}.
$$
Let us compute this quantity to see its $t$ dependence. We first bring all $\tilde X_+$ components to the left untill they hit to the boundary state $\langle N|$. After all of the right moving components are reflected into the left moving ones $X_-$, the state $\langle N|$ will be turned into the vacuum state $\langle 0|$. Hence the rest we have to calculate is 
$$
{\langle 0|e^{i\sum_{j=1}^{M+2}k_jX_-(\bar z_j^{-1})}e^{-i\theta_a J^a}e^{i\sum_{j=1}^{M+2}k_jX_+(z_j)}|0\rangle
\over
\langle 0|e^{i\sum_{j=1}^{M}k_jX_-(\bar z_j^{-1})}e^{-i\theta_a J^a}e^{i\sum_{j=1}^{M}k_jX_+(z_j)}|0\rangle},
$$
where we used the notations
$$
k_{M+1}=-k_{M+2}={1\over \sqrt2},\qquad z_{M+1}=z,\quad z_{M+2}=z'.
$$
Applying the formula (\ref{J^theta}), we obtain
\begin{eqnarray*}
W&=&\left({(\bar z^{-1}-z)(\bar z'^{-1}-z')\over(\bar z'^{-1}-z)(\bar z^{-1}-z')}\right)^{{1\over 2}\cos(2\pi|g|)}\\
&&\times \prod_{j=1}^M\left({(\bar z^{-1}-z_j)(\bar z_j^{-1}-z)\over (\bar z'^{-1}-z_j)(\bar z_j^{-1}-z')}\right)^{{1\over\sqrt2}k_j\cos(2\pi|g|)}\\
\end{eqnarray*}
Under the conditions that all $z_j$'s are on the unit circle and $\sum_jk_j=0$ we can write the second factor of $W$ as
$$
e^{\Xi(t,\tilde t,z,z')\cos(2\pi|g|)}.
$$
If we combine these results together the modification to the $\tau$ function due to the boundary is rather simple
\begin{eqnarray*}
&&\tau^B_{1sol}=1+a{y^{\cos^2(\pi|g|)}\over\sqrt{(z-z')(\bar z-\bar z')y}}e^{\Xi(t,\tilde t, z, z')2\cos^2(\pi|g|)},
\end{eqnarray*}
where $y$ is the cross ratio defined by
$$
y:={(\bar z^{-1}-z)(\bar z'^{-1}-z')\over(\bar z'^{-1}-z)(\bar z^{-1}-z')}.
$$

Let us see how the profile of the soliton solution of the Toda equation will be affected by a change of the boundary parameter $g$. Comparing with the data (\ref{Toda velocity}) we find
$$
e^{r_m}-1={\omega'^2\over \cosh^2(\kappa'm+\omega' t+\delta')}
$$
where
$$
\kappa'=2\sqrt 2\kappa\cos^2(\pi|g|),\ \ \omega'=2\sqrt2\omega\cos^2(\pi|g|),
$$
$$
\delta'=\delta +{1\over 4}\cos(2\pi|g|)\ln y
$$
and assumed $z$ and $z'$ being real.
$$
t:={t_3+\tilde t_3\over 2},\qquad m:={t_1+\tilde t_1\over 2}
$$
An interesting feature of this result is that the soliton is completely frozen when $|g|$ is a half integer corresponding to the Dirichlet boundary condition in the string picture. We have also calculated two soliton solution influenced by the boundary and found that they are stable under the collisions. Details will be reported elsewhere.

\noindent
{\bf Acknowledgement}

The authors would like to thank Dr. N.Kitazawa for discussions. This work is partially supported by TMU fund.


\begin{thebibliography}{99}
\bibitem{CKLM}
C.Callan, I.Klebanov, A.Ludwig and J.Maldacena, Nucl.Phys. B{\bf 422} (1994) 417-448 
\bibitem{SS} S.~Saito, String theories and Hirota's bilinear difference equation, Phys. Rev. Lett. {\bf 59} (1987) 1798-1801, 

S.~Saito, String vertex on the Grassmann manifold, Phys. Rev. D{\bf 37} (1988) 990-995.

K.~Sogo, A way from string to soliton - Introduction of KP coordinate to string amplitudes-, J. Phys. Soc. Jpn. {\bf 56} (1987) 2291-2297.
\bibitem{Hirota}
R.~Hirota, Discrete analogue of a generalized Toda equation, J.Phys. Soc. Jpn. {\bf 50} (1981) 3785-3791.
\bibitem{Miwa}
T.~Miwa, On Hirota's difference equations, Proc. Japan Acad. A {\bf 58} (1982) 9-12.

\bibitem{Sato}
M.Sato and Y.Sato, Publ. RIMS {\bf 388} (1980) 183; {\it ibid} {\bf 414} (1981) 181; M.Sato, {\it ibid}. {\bf 433} (1981) 30.
\bibitem{KS}
H.Kato and S.Saito, Lett. Math. Phys. {\bf 18} (1989) 177-183
Generalization of boson-fermion equivalence and Fay's addition theorem
\bibitem{Mumford}
D.~Mumford, {\it Tata lectures on theta I, II} (Birkh\"auser, 1983).
\bibitem{Fay}
J.D.Fay, "Theta functions on Riemann surfaces", Lect. Notes in Math. 352 (Springe-Verlag, 1973)

\bibitem{DJKM} E.~Date, M.~Jimbo, M.~Kashiwara and T.~Miwa, Operator approach to the Kadomtsev-Petviashvili equation - Transformation groups for soliton equations III, VI, J. Phys. Soc. Jpn. {\bf 50} (1981) 3806-3818.

\bibitem{SS2}
N.Saitoh and S.Saito, J. Physics A: Math. Gen. {\bf 23} (1990) 3017-3027

Integrable lattice gauge model based on soliton theory

\end{thebibliography}
\end{document}